\documentclass[a4paper,11pt]{article}
\usepackage{pos}

\usepackage{xspace} % To avoid problems with missing or double spaces after
\usepackage{upgreek} % Adds in support for greek letters in roman typeset
\def\PB {\ensuremath{\mathrm{B}}\xspace}

\def\PD {\ensuremath{\mathrm{D}}\xspace}

\def\PJ {\ensuremath{\mathrm{J}}\xspace}
\def\PK {\ensuremath{\mathrm{K}}\xspace}

\def\PM {\ensuremath{\mathrm{M}}\xspace}

\def\PP {\ensuremath{\mathrm{P}}\xspace}

\def\PS {\ensuremath{\mathrm{S}}\xspace}

\def\PX {\ensuremath{\mathrm{X}}\xspace}

\def\Pb {\ensuremath{\mathrm{b}}\xspace}
\def\Pc {\ensuremath{\mathrm{c}}\xspace}
\def\Pd {\ensuremath{\mathrm{d}}\xspace}

\def\Pf {\ensuremath{\mathrm{f}}\xspace}

\def\Pi {\ensuremath{\mathrm{i}}\xspace}

\def\Pp {\ensuremath{\mathrm{p}}\xspace}

\def\Pu {\ensuremath{\mathrm{u}}\xspace}

\def\uquark {\ensuremath{\Pu}\xspace}
\def\dquark {\ensuremath{\Pd}\xspace}
\def\bquark {\ensuremath{\Pb}\xspace}

\def\cquark {\ensuremath{\Pc}\xspace}

\def\B {{\ensuremath{\PB}}\xspace}
\def\POmega {\ensuremath{\Omega}\xspace}
\def\PLambda {\ensuremath{\Lambda}\xspace}
\def\PXi {\ensuremath{\Xi}\xspace}
\def\Ppi {\ensuremath{\uppi}\xspace}
\def\Ppsi {\ensuremath{\uppsi}\xspace}
\def\Pmu {\ensuremath{\upmu}\xspace}
\def\Pnu {\ensuremath{\upnu}\xspace}
\def\PSigma {\ensuremath{\Sigma}\xspace}
\def\proton {\ensuremath{\Pp}\xspace}
\def\kaon {{\ensuremath{\PK}}\xspace}
\def\pion {{\ensuremath{\Ppi}}\xspace}
\def\p {\ensuremath{\Pp}\xspace}
\def\muon {{\ensuremath{\Pmu}}\xspace}
\def\pim {{\ensuremath{\pion^-}}\xspace}
\def\pip {{\ensuremath{\pion^+}}\xspace}
\def\pipm {{\ensuremath{\pion^\pm}}\xspace}
\def\Km {{\ensuremath{\kaon^-}}\xspace}
\def\Kp {{\ensuremath{\kaon^+}}\xspace}
\def\Xibz {{\ensuremath{\PXi^0_\bquark}}\xspace}
\def\Xicp {{\ensuremath{\PXi^+_\cquark}}\xspace}
\def\Dz {{\ensuremath{\PD^0}}\xspace}
\def\Dp {{\ensuremath{\PD^+}}\xspace}
\def\Bd {{\ensuremath{\PB^0}}\xspace}
\def\Bu {{\ensuremath{\B^+}}\xspace}
\def\Bub {{\ensuremath{\B^-}}\xspace}
\def\Bcm {{\ensuremath{\B_\cquark^-}}\xspace}
\def\Bcp {{\ensuremath{\B_\cquark^+}}\xspace}
\def\Ocz {{\ensuremath{\POmega^0_\cquark}}\xspace}
\def\Obm {{\ensuremath{\POmega^-_\bquark}}\xspace}
\def\ObmA {{\ensuremath{\POmega_\bquark(6316)^-}}\xspace}
\def\ObmB {{\ensuremath{\POmega_\bquark(6330)^-}}\xspace}
\def\ObmC {{\ensuremath{\POmega_\bquark(6340)^-}}\xspace}
\def\ObmD {{\ensuremath{\POmega_\bquark(6350)^-}}\xspace}
\def\Lz {{\ensuremath{\PLambda}}\xspace}
\def\Lb{{\ensuremath{\PLambda_\bquark^0}}\xspace}
\def\Lcp{{\ensuremath{\PLambda_\cquark^+}}\xspace}
\def\Lboldone{{\ensuremath{\Lz_\bquark(5912)^0}}\xspace}
\def\Lboldtwo{{\ensuremath{\Lz_\bquark(5920)^0}}\xspace}
\def\Lbnewone{{\ensuremath{\Lz_\bquark(6146)^0}}\xspace}
\def\Lbnewtwo{{\ensuremath{\Lz_\bquark(6152)^0}}\xspace}
\def\Lboned{{\ensuremath{\Lz_\bquark(\rm 1D)^0}}\xspace}
\def\Lbstarr{{\ensuremath{\Lz_\bquark^{**0}}}\xspace}
\def\Lbtwos{{\ensuremath{\Lz_\bquark(2\PS)^0}}\xspace}
\def\jpsi {{\ensuremath{{\PJ\mskip -3mu/\mskip -2mu\Ppsi\mskip 2mu}}}\xspace}

\def\mum {{\ensuremath{\Pmu^-}}\xspace}

\def\numb {{\ensuremath{{\overline \Pnu}_{\muon}}}\xspace}
\def\Lbpipi {{\ensuremath{\Lb\pip\pim}}\xspace}
\def\Lbpipinosign {{\ensuremath{\Lb\pion\pion}}\xspace}
\def\Lbpipiws {{\ensuremath{\Lb\pipm\pipm}}\xspace}
\def\Lbpipm {{\ensuremath{\Lb\pipm}}\xspace}
\def\Lbpippip {{\ensuremath{\Lb\pip\pip}}\xspace}

\def\Lcppim {{\ensuremath{\Lcp\pim}}\xspace}
\def\JpsipKm {{\ensuremath{\jpsi\p\Km}}\xspace}
\def\Sigmabmp {{\ensuremath{\PSigma_\bquark^{\mp}}}\xspace}
\def\Sigmabsmp {{\ensuremath{\PSigma_\bquark^{*\mp}}}\xspace}
\def\Sigmabsspm {{\ensuremath{\PSigma_\bquark^{(*)\pm}}}\xspace}
\def\Sigmabsmppipm {{\ensuremath{\PSigma_\bquark^{(*)\mp}\pi^\pm}}\xspace}

\def\lhc {\mbox{LHC}\xspace}
\def\lhcb {\mbox{LHCb}\xspace}

\def\JP {\ensuremath{\PJ^\PP}\xspace}
\def\pp {\ensuremath{\proton\proton}\xspace}
\def\fu {\ensuremath{\Pf_{\uquark}}\xspace}
\def\fd {\ensuremath{\Pf_{\dquark}}\xspace}
\def\fc {\ensuremath{\Pf_{\cquark}}\xspace}
\def\fcratio {\ensuremath{\dfrac{\fc}{\fu+\fd}}\xspace}
\def\X {\ensuremath{\PX}\xspace}

\newcommand{\aunit}[1]{\ensuremath{\text{\,#1}}}
\newcommand{\tev}{\aunit{Te\kern -0.1em V}\xspace}
\newcommand{\mev}{\aunit{Me\kern -0.1em V}\xspace}
\newcommand{\gev}{\aunit{Ge\kern -0.1em V}\xspace}
\def\fb {\ensuremath{\aunit{fb}}\xspace}
\def\invfb {\ensuremath{\fb^{-1}}\xspace}
\def\M {\ensuremath{\PM}\xspace}
\def\pt {\ensuremath{p_{\mathrm{T}}}\xspace}
\def\BcmToJpsiMumNumb {{\Bcm\to\jpsi\mum\numb}\xspace}
\def\BdToDpXMumNumb {{\Bd\to\Dp\X\mum\numb}\xspace}
\def\BudToDzXMumNumb {{\Bub\to\Dz\X\mum\numb}\xspace}

\makeatletter
\g@addto@macro\bfseries{\boldmath}
\makeatother

\title{Studies of b-hadrons at $\lhcb$}
\author*[a,1]{Viacheslav Matiunin}

\affiliation[a]{Institute for Theoretical and Experimental Physics, NRC <<Kurchatov Institute>>,\\
	B. Cheremushkinskaya st. 25, Moscow, 117218, Russia}

\note{On behalf of the $\lhcb$ Collaboration.}

\emailAdd{viacheslav.matiunin@cern.ch}

\abstract{
	A large data set collected at the LHCb experiment in proton-proton collisions during Runs~1 and~2 of the Large Hadron Collider has opened a unique possibility to study heavy beauty hadron states and to broaden knowledge of their spectroscopy and production.
	Recent results on searches for new excited \mbox{$\bquark$-hadron} states, and studies of \mbox{$\bquark$-hadron} production, are reviewed.
	In particular,
	the observation of new excited $\Obm$~states,
	the observation of two new narrow $\Lbnewone$ and $\Lbnewtwo$~states,
	the observation of new $\Lbstarr$ state consistent with the $\Lbtwos$~prediction,
	a precise measurement of the mass and width of $\Lboldone$ and $\Lboldtwo$~states,
	and
	a measurement of the $\Bcm$~meson production fraction and $\Bcm-\Bcp$ production asymmetry in 7~and 13~$\tev$ of proton-proton collisions
	are presented.
}

\FullConference{%
  40th International Conference on High Energy Physics - ICHEP2020\\
  July 28 - August 6, 2020\\
  Prague, Czech Republic (virtual meeting)
}

\begin{document}
\maketitle

\section{Introduction}
Over the last few years numerous outstanding results have been obtained within the analyses of \mbox{\bquark-hadrons}.
Among those there are observation of pentaquark and tetraquark resonances, observation of series of excited heavy baryons and many others.
Still, many of the conventional states remain unobserved, parameters of some of the known hadron states are poorly measured and on top of that there are a number of the states which do not fit into the conventional quarkonium spectrum.
Hence study of \mbox{\bquark-hadrons} spectroscopy is of particular interest in the modern high energy physics.

The results described below are based on the data samples collected by the \lhcb experiment in proton-proton~($\pp$) collision at the Large Hadron Collider~(LHC) with centre-of-mass energies $\sqrt{\rm s}=7$ and $8\,\tev$ corresponding to a total integrated luminosity of $3\,\invfb$ (\lhc Run~1) and with centre-of-mass energy $\sqrt{\rm s}=13\,\tev$ corresponding to a total integrated luminosity of $6\,\invfb$ (\lhc Run~2).

\section{Observation of excited $\Obm$~states}
\begin{figure}[b]
	\centering
	\includegraphics[height=4cm,width=7.5cm]{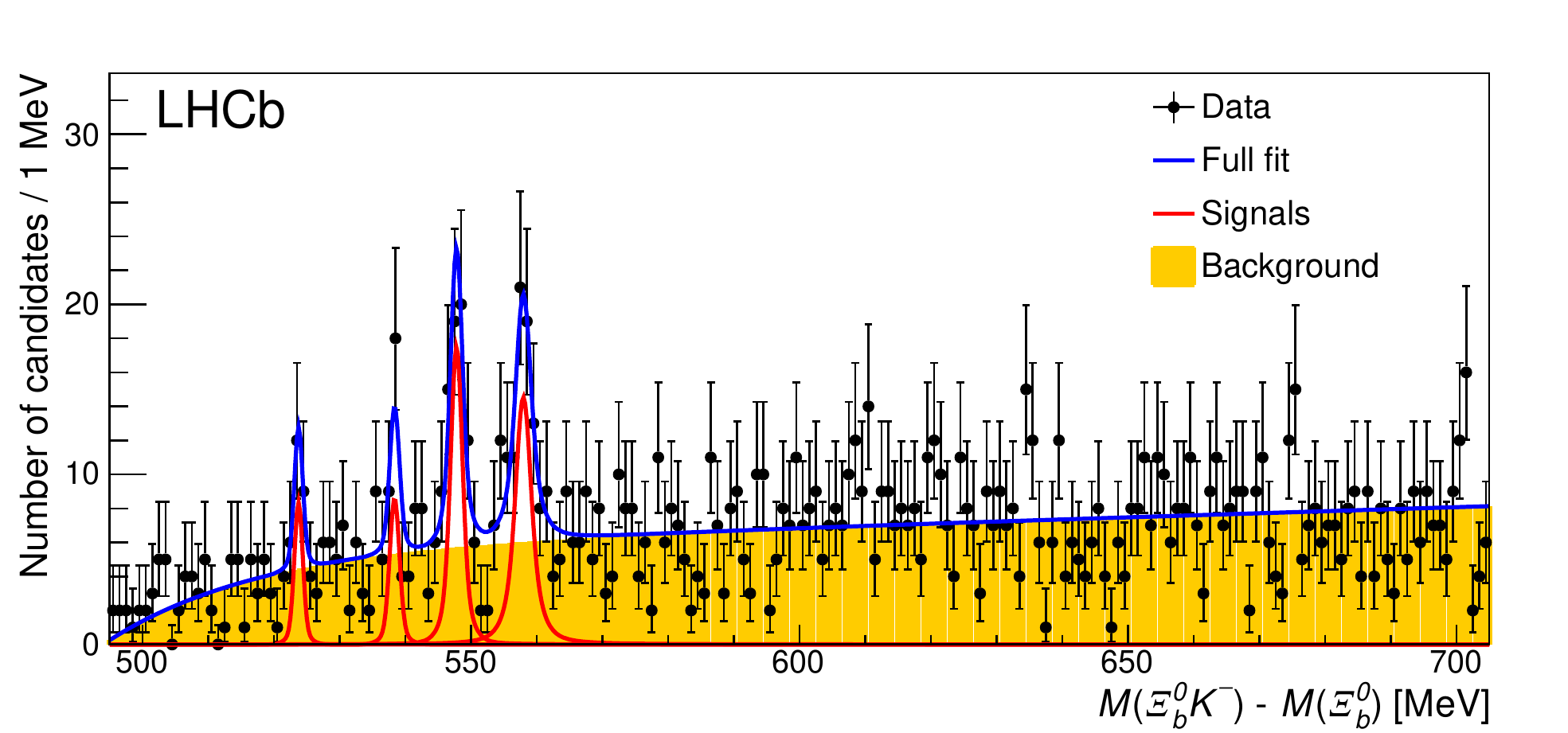}
	\put(-190,80){(a)}
	\includegraphics[height=4cm,width=7.5cm]{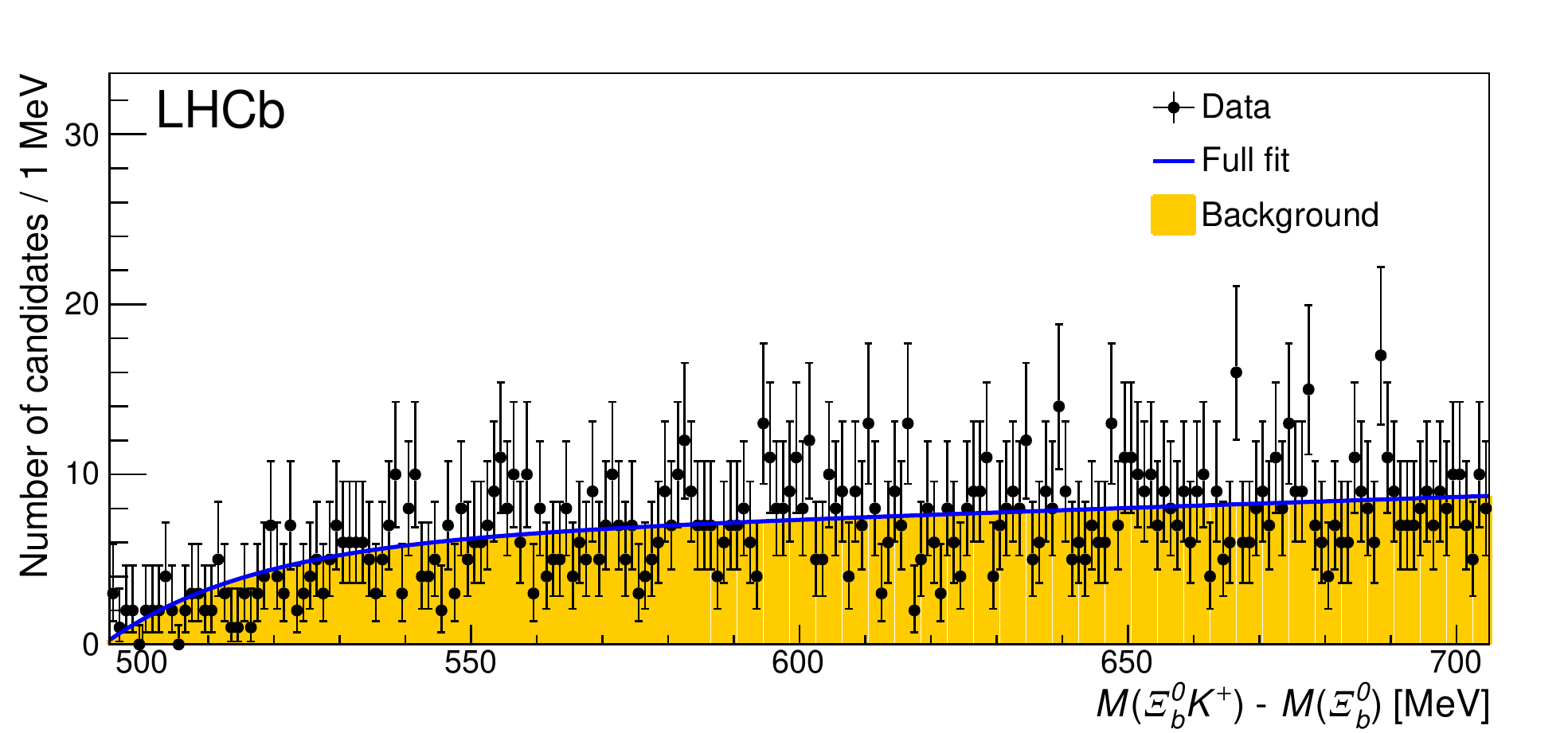}
	\put(-190,80){(b)}
	\caption{\small
		The distributions of (a)~right-sign $\M(\Xibz\Km)-\M(\Xibz)$ and (b)~wrong-sign $\M(\Xibz\Kp)-\M(\Xibz)$ mass differences.
		Different components employed in the fit are indicated in the legend.
	}
	\label{Fig1}
\end{figure}

Recently, the new five narrow excited $\Ocz$~baryons has been observed~\cite{LHCb-PAPER-2017-002}.
Some of the theoretical models describing $\Ocz$~peaks also predict $\Obm$~states decaying to $\Xibz\Km$~final state.
Therefore, it is of great interest to search for the analogous excited states with the \lhcb~experiment data.
The $\Xibz\Km$~mass spectrum is studied to search for narrow resonances close to the kinematic threshold~\cite{LHCb-PAPER-2019-042}.
The analysis is done using $\pp$~collision data samples corresponding to an integrated luminosity of $9.0~\invfb$~(\lhc Run~1 and 2).
The $\Xibz$~baryons are reconstructed using $\Xicp\pim$~decay mode, where $\Xicp$~baryons are reconstructed with the $\p\Km\pip$~final state.
The distributions of right-sign combination mass difference $\M(\Xibz\Km)-\M(\Xibz)$ and wrong-sign combination mass difference $\M(\Xibz\Kp)-\M(\Xibz)$ are shown in Fig.~\ref{Fig1}\,(a) and~(b), respectively.
Four narrow peaks are seen in right-sign distribution.
Simultaneous fit to wrong-sign and right-sign spectra with common background shape is performed.
The peak parameters obtained from the fit results together with the significance of the peaks are shown in Table~\ref{Table1}.
The mass and width of the new resonances are consistent with expectations for excited $\Obm$~states.

\begin{table*}[t]
	\begin{center}
		\caption{\small
			The masses, 90\% (95\%) confidence level upper limits on the natural widths and global~(local) significance for the four peaks.
			The uncertainties are statistical, systematic, and due to the world-average value of the $\Xibz$ mass (for the masses).
			For the $\ObmD$ peak, the central value of the width is also indicated.}
		\begin{tabular}{lccc}
			\hline
			& Mass [Me\kern -0.1em V] & Width [Me\kern -0.1em V] & Significance [$\sigma$] \\
			\hline\\[-5mm]
			$\ObmA$ & $6315.64\pm0.31\pm0.07\pm0.50$  & $<2.8~(4.2)$ & 2.1~(3.6)\\[-1mm]
			$\ObmB$ & $6330.30\pm0.28\pm0.07\pm0.50$  & $<3.1~(4.7)$ & 2.6~(3.7)\\[-1mm]
			$\ObmC$ & $6339.71\pm0.26\pm0.05\pm0.50$  &  $<1.5~(1.8)$ & 6.7~(7.2)\\[-1mm]
			$\ObmD$ & $6349.88\pm0.35\pm0.05\pm0.50$  & $<2.8~(3.2)$ & 6.2~(7.0)\\[-1mm]
			&& $~~\,\,1.4\,^{+\,1.0}_{-\,0.8}\pm0.1$  \\[0.5mm]
			\hline
		\end{tabular}
		\label{Table1}
	\end{center}
\end{table*}

\section{Excited $\Lb$~baryons}
Beyond the lightest beauty baryon, $\Lb$~baryon, a rich spectrum of radially and orbitally excited states is expected at higher masses.
The excited states has been searched in the $\Lbpipi$~spectrum by the LHCb experiment with the discovery of two narrow states~\cite{LHCb-PAPER-2012-012}.
Therefore, it is both interesting and promising to study the $\Lbpipi$~spectrum with the full statistics collected by the LHCb experiment.
The $\Lbpipi$~mass spectrum is studied by the LHCb experiment using the full \lhc Run~1 and~2 data sample of $\pp$~collisions, corresponding to an integrated luminosity of 9~$\invfb$~\cite{LHCb-PAPER-2019-025,LHCb-PAPER-2019-045}.
The $\Lb$~baryon is reconstructed using two decay modes $\Lcppim$ and $\JpsipKm$.
The study is performed in three $\Lbpipi$~mass intervals: high-, middle- and low-mass.
\begin{figure}[b]
	\centering
	\includegraphics[width=5cm,height=5.8cm,trim={1.35cm 1cm 0 0},clip]{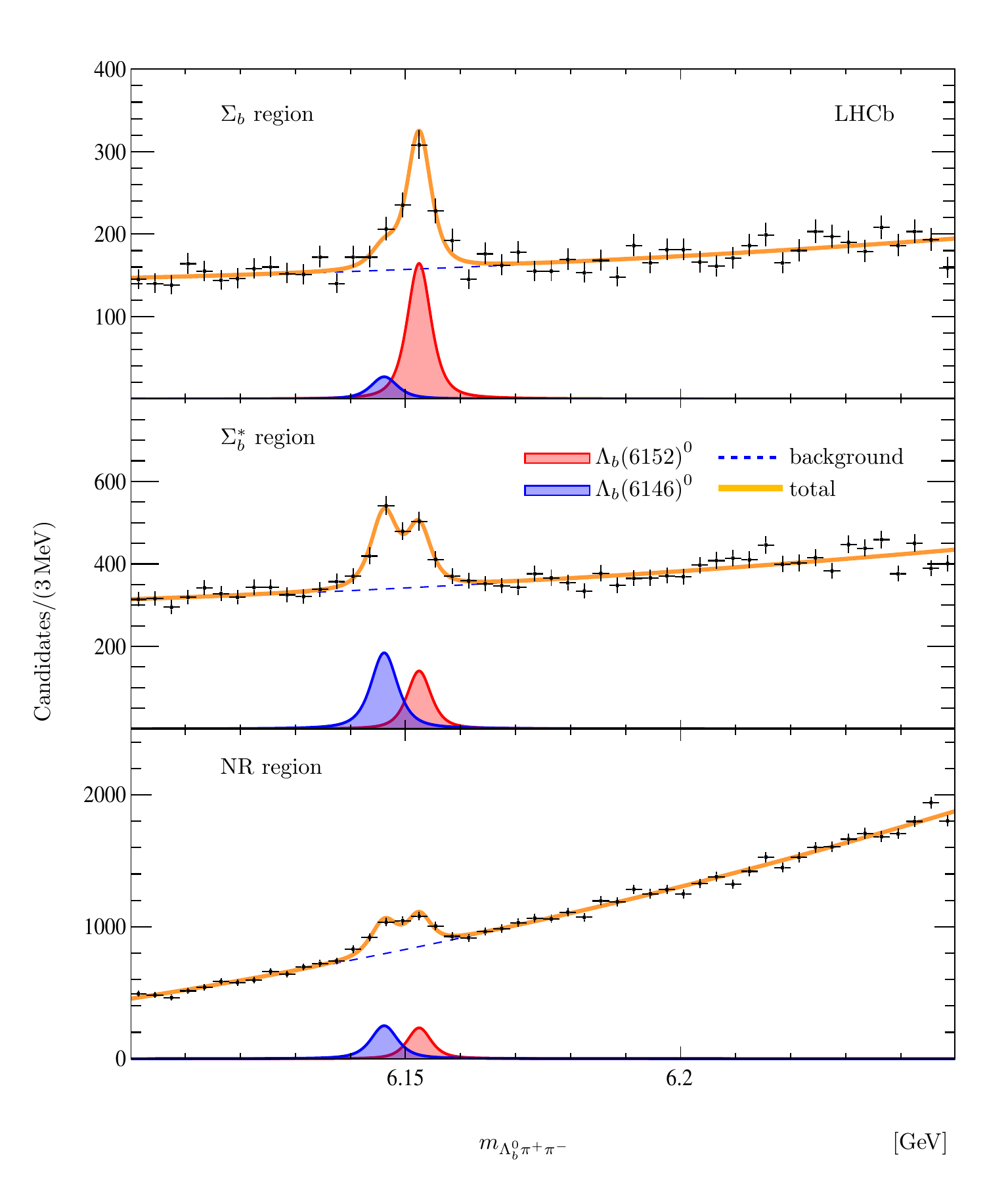}
	\put(-42,144.5){(a)}
	\put(-148,100){\scriptsize\rotatebox[]{90}{Candidates / $3\,\mev$}}
	\put(-84,-3){\scriptsize$\M(\Lbpipi)\quad\quad[\gev]$}
	\includegraphics[width=5cm,height=5.8cm,trim={1.35cm 1cm 0 0},clip]{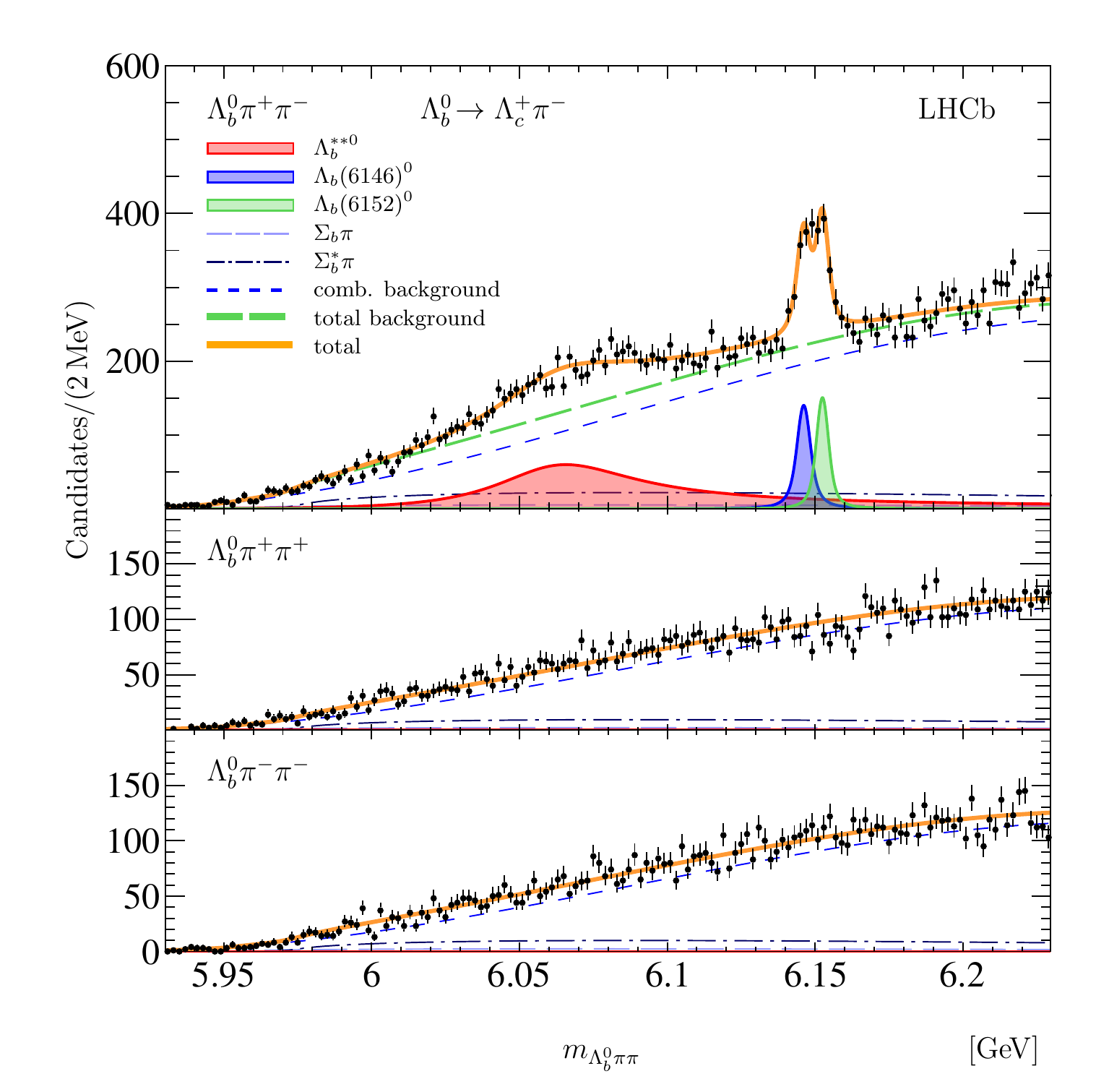}
	\put(-42,144.5){(b)}
	\put(-148,100){\scriptsize\rotatebox[]{90}{Candidates / $0.4\,\mev$}}
	\put(-83,-3){\scriptsize$\M(\Lbpipinosign)\quad\quad\quad[\gev]$}
	\includegraphics[width=5cm,height=5.8cm,trim={1.35cm 1cm 0 0},clip]{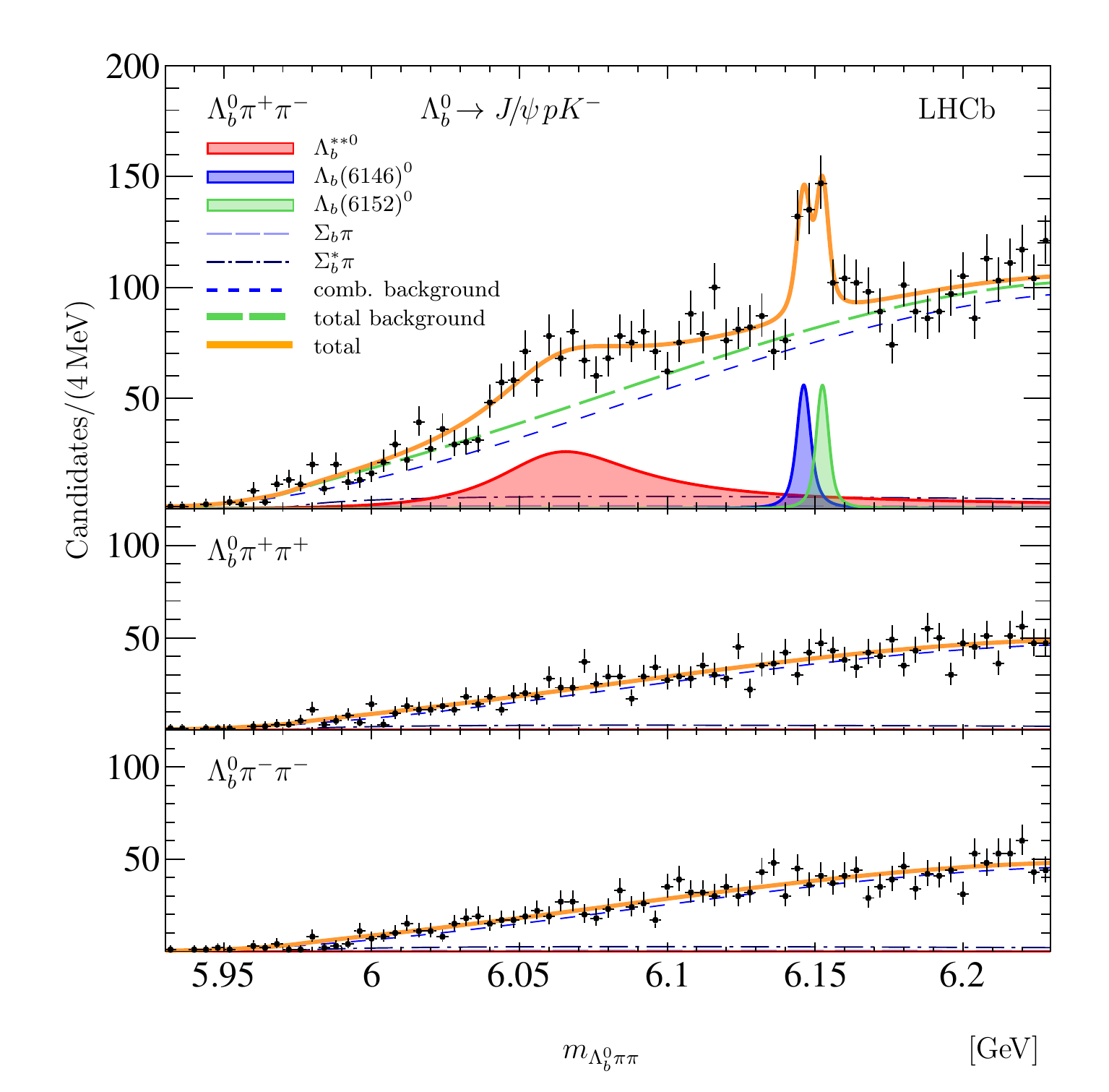}
	\put(-42,144.5){(c)}
	\put(-148,100){\scriptsize\rotatebox[]{90}{Candidates / $0.4\,\mev$}}
	\put(-83,-3){\scriptsize$\M(\Lbpipinosign)\quad\quad\quad[\gev]$}
	\caption{\small
		(a)~Mass distributions of $\Lbpipi$~combination for the three regions in $\Lbpipm$ mass: (top)~$\Sigmabmp$, (middle)~$\Sigmabsmp$ and (bottom)~nonresonant region.
		Mass spectra of (top)~$\Lbpipi$, (middle)~$\Lbpippip$ and (bottom)~$\Lbpippip$~combinations for the (b)~$\Lb\to\Lcppim$ and (c)~$\Lb\to\JpsipKm$~sample.
		Different components employed in the fit are indicated in the legend.
	}
	\label{Fig2}
\end{figure}
In the high $\Lbpipi$~mass interval a new peaking structure at approximately $6.15\,\gev$ is observed~\cite{LHCb-PAPER-2019-025}.
The peak is above $\Sigmabsmppipm$~threshold, hence, spectrum is investigated in three non-overlapping $\Lbpipm$~mass regions: two resonant $\Sigmabmp$ and $\Sigmabsmp$ regions and nonresonant one.
Simultaneous fit to the three $\Lbpipi$~mass distributions is performed.
The two-peak hypothesis is favoured with respect to the single-peak hypothesis with more than $7\,\sigma$~statistical significance.
The $\Lbpipi$~mass distributions are shown in Fig.~\ref{Fig2}\,(a).
Mass and width of the two peaks are
%\begin{linenomath*}
	\begin{align}
		\M_{\Lbnewone} = 6146.17 \pm 0.33 \pm 0.22 \pm 0.16 \mev, \quad \Gamma_{\Lbnewone} = 2.9 \pm 1.3 \pm 0.3\mev,\\
		\M_{\Lbnewtwo} = 6152.51 \pm 0.26 \pm 0.22 \pm 0.16 \mev, \quad \Gamma_{\Lbnewtwo} = 2.1 \pm 0.8 \pm 0.3\mev,
	\end{align}
%\end{linenomath*}
\noindent here and throughout the Section the first uncertainty is statistical, the second one is systematic and third one is due to the uncertainty in the nominal $\Lb$~baryon mass.
The measured masses are consistent with predictions for \Lboned~doublet with $\JP=\frac{3}{2}^+$ and~$\frac{5}{2}^+$~\cite{Chen:2014nyo,PhysRevD.34.2809}.

\begin{figure}[b]
	\centering
	\includegraphics[width=5.7cm,height=5.8cm,trim={1.35cm 1cm 0 0},clip]{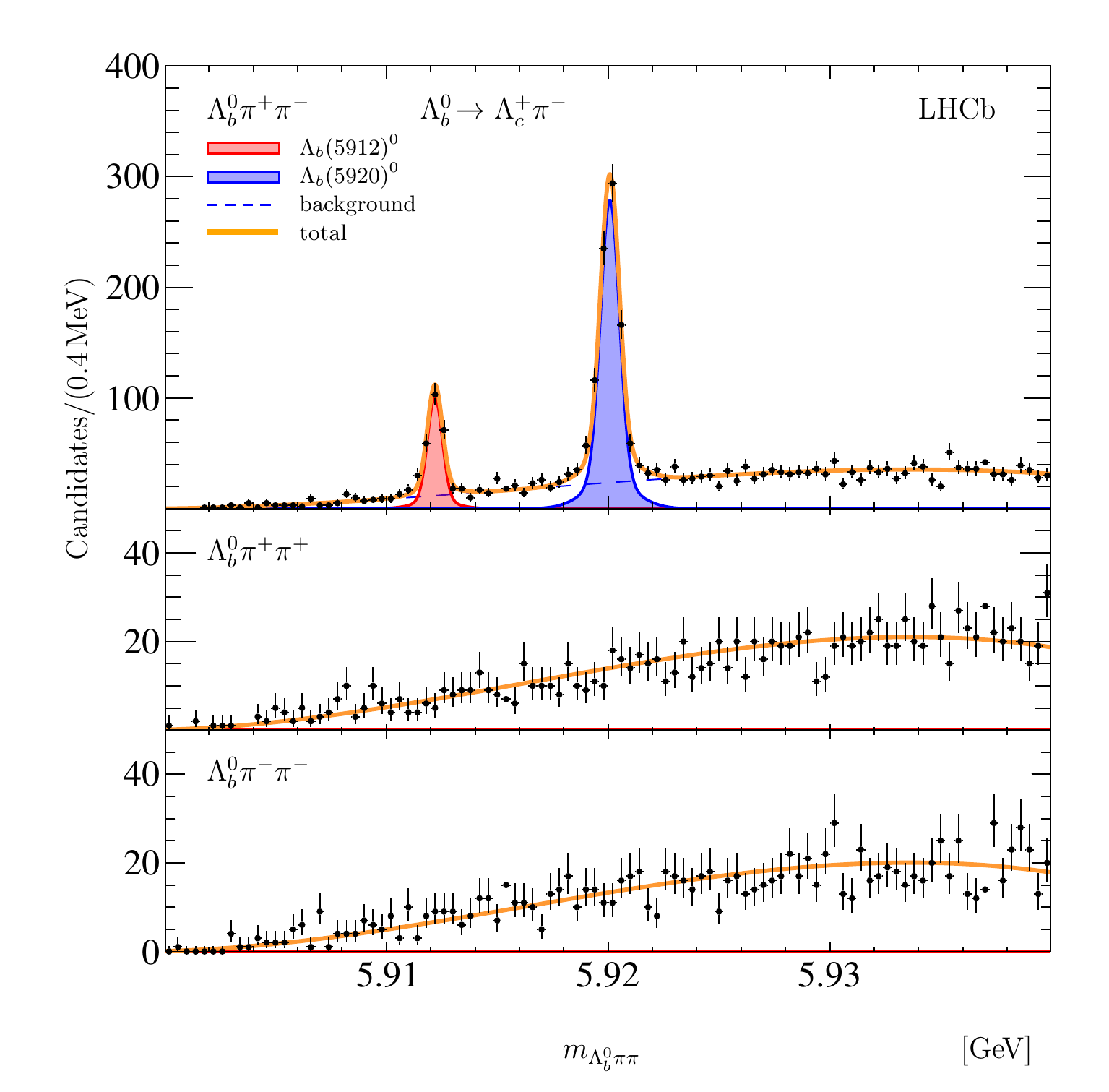}
	\put(-50,143){(a)}
	\put(-170,100){\scriptsize\rotatebox[]{90}{Candidates / $0.4\,\mev$}}
	\put(-90,-3){\scriptsize$\M(\Lbpipinosign)\;\;\,\quad\quad\quad[\gev]$}
	\hspace*{0.5cm}
	\includegraphics[width=5.7cm,height=5.8cm,trim={1.35cm 1cm 0 0},clip]{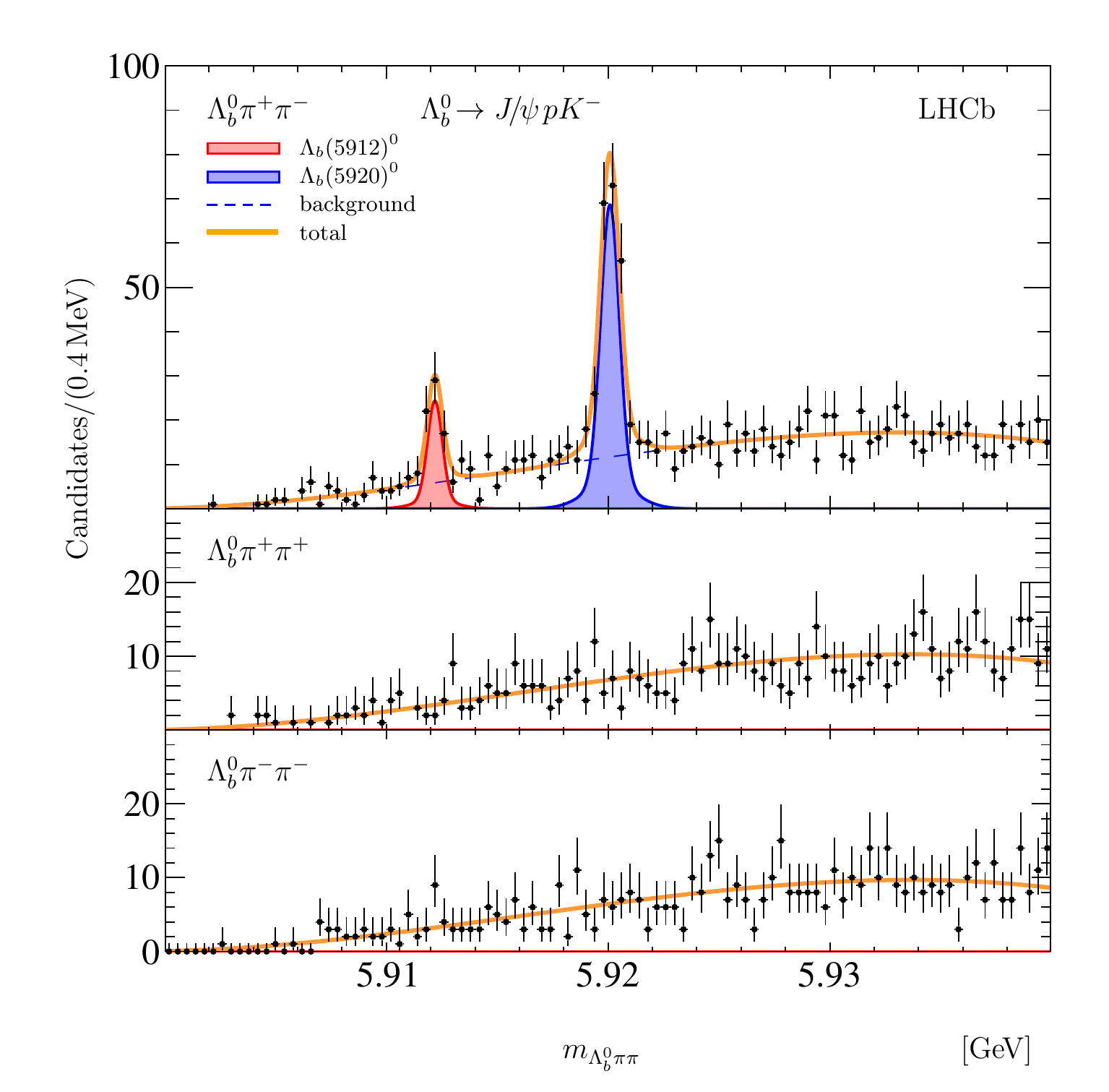}
	\put(-50,143){(b)}
	\put(-170,100){\scriptsize\rotatebox[]{90}{Candidates / $0.4\,\mev$}}
	\put(-90,-3){\scriptsize$\M(\Lbpipinosign)\;\;\,\quad\quad\quad[\gev]$}
	\caption{\small
		Mass spectra of (top)~$\Lbpipi$, (middle)~$\Lbpippip$ and (bottom)~$\Lbpippip$~combinations for the (a)~$\Lb\to\Lcppim$ and (b)~$\Lb\to\JpsipKm$~sample.
		Different components employed in the fit are indicated in the legend.
	}
	\label{Fig3}
\end{figure}

In the intermediate $\Lbpipi$~mass interval a new baryon state is observed~\cite{LHCb-PAPER-2019-045}.
The simultaneous fit to the six mass distributions where $\Lb$~baryon is reconstructed via two different decay modes and in each case for one right-sign $\Lbpipi$ and two wrong-sign $\Lbpipiws$ spectra is performed.
The $\Lbpipi$~mass distributions are shown in Fig.~\ref{Fig2}\,(b) and~(c).
Mass and width of the new peak are
\begin{align}
\M_{\Lbstarr} = 6072.3 \pm 2.9 \pm 0.6 \pm 0.2 \mev, \quad \Gamma_{\Lbstarr} = 72 \pm 11 \pm 2\mev.
\end{align}
The observed peak is consistent with broad excess reported by the CMS experiment~\cite{Sirunyan:2020gtz}.
The measured mass and width are in agreement with expectations for the \Lbtwos~state.
The contributions from $\Sigmabsspm$~resonances are also studied and the nonresonant component is found to give a dominant contribution.

In the low $\Lbpipi$~mass interval the two states observed earlier~\cite{LHCb-PAPER-2012-012} are confirmed.
The simultaneous fit to the six mass distributions where $\Lb$~baryon is reconstructed via two different decay modes and in each case for one right-sign $\Lbpipi$ and two wrong-sign $\Lbpipiws$ spectra is performed.
The mass spectra are shown in Fig.~\ref{Fig3}.
The mass and width of the states are measured to be
\vspace*{-3mm}
\begin{align}
		\M_{\Lboldone} &= 5912.21 \pm 0.03 \pm 0.01 \pm 0.21 \mev, \quad \Gamma_{\Lboldone} < 0.25\,(0.28)\mev,\\
		\M_{\Lboldtwo} &= 5920.11 \pm 0.02 \pm 0.01 \pm 0.21 \mev, \quad \Gamma_{\Lboldtwo} < 0.19\,(0.20)\mev.
\end{align}
\noindent For the natural widths the upper limits at $90\%~(95\%)$ confidence level is specified.
The parameters are measured with about four times higher precision with respect to those reported in Ref.~\cite{LHCb-PAPER-2012-012}.

\section{Measurement of the $\Bcm$~meson production fraction and asymmetry in $\pp$~collisions at 7~and 13~$\tev$}
The $\bquark$-hadrons cross-sections as functions of transverse momentum~($\pt$) and pseudorapidity~($\eta$) are predicted using non-relativistic quantum chromodynamics along with fragmentation functions.
The corresponding literature is systematically reviewed in the Ref.~\cite{Brambilla:2010cs}.
The measurement of the $\Bcm$~production fraction would allow to further probe quantum chromodynamics.

\begin{figure}[b]
	\centering
	\includegraphics[height=4cm,trim={4cm 3.5cm 4cm 4cm},clip]{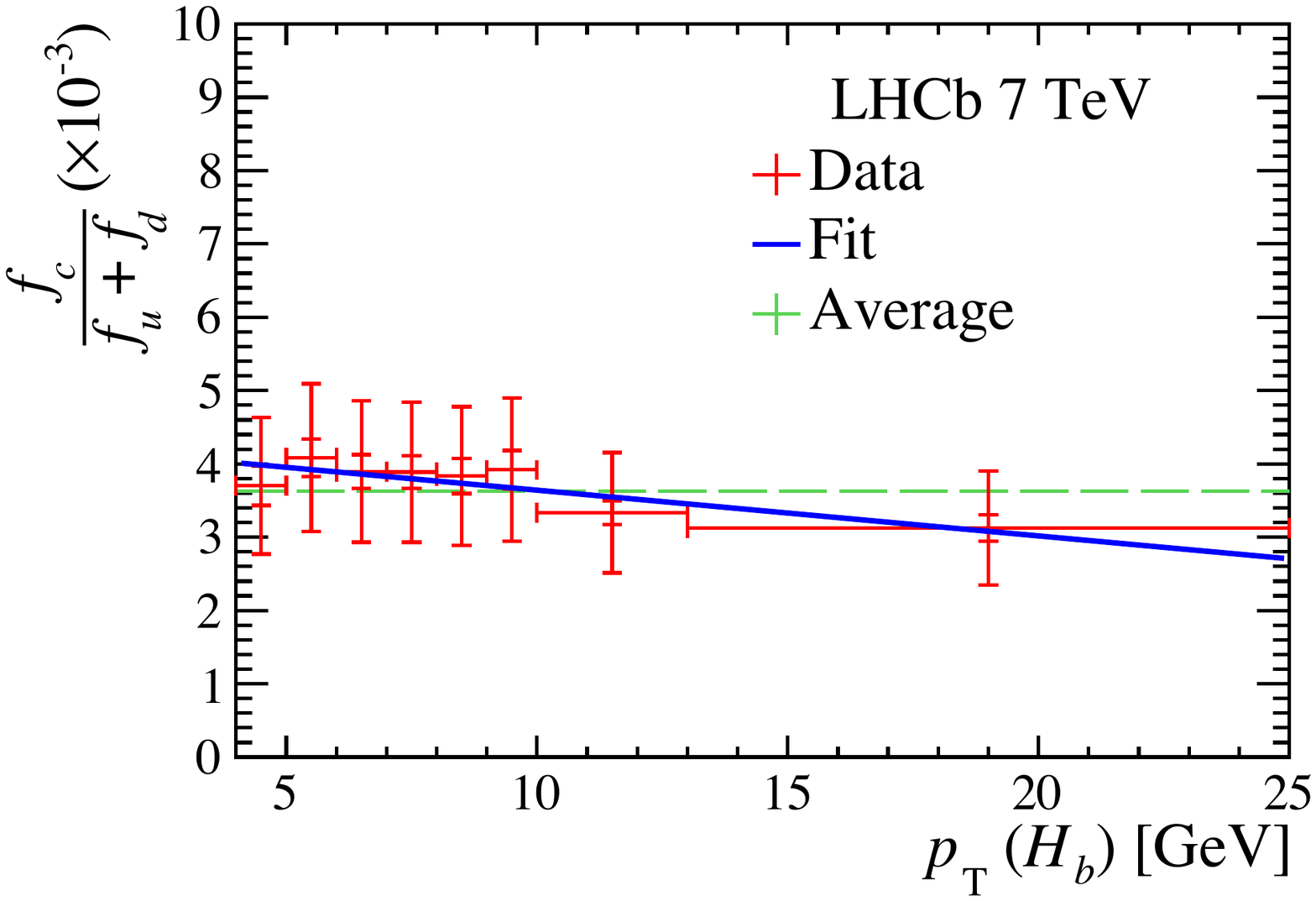}
	\put(-114,83){(a)}
	\hspace*{0.5cm}
	\includegraphics[height=4cm,trim={4cm 3.5cm 4cm 4cm},clip]{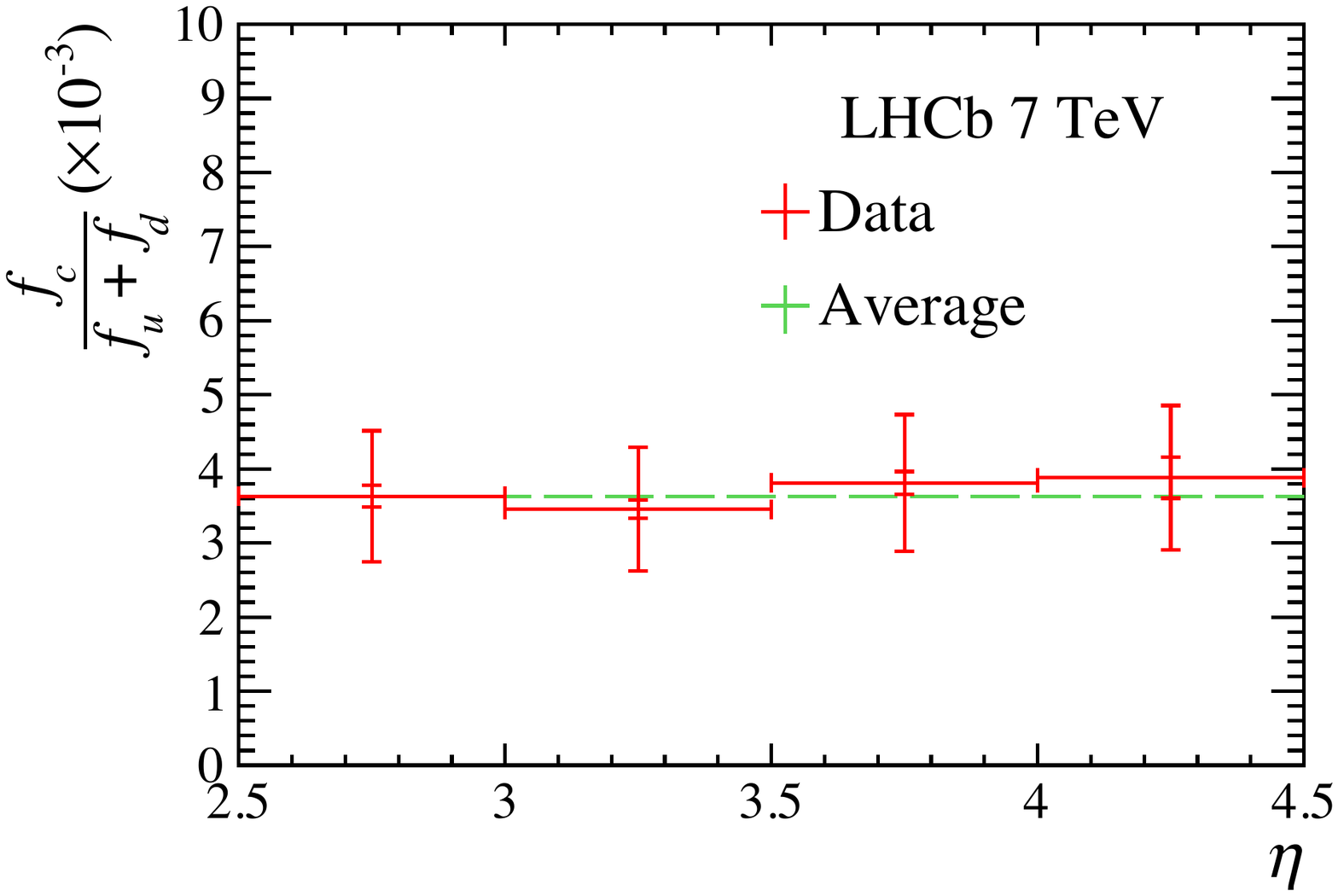}
	\put(-114,83){(b)}
	\caption{\small
		Ratio of production fractions as a function of (a)~$\pt$ and (b)~$\eta$ of $\B$~mesons for the 7~$\tev$ data sample.
		The statistical uncertainties are shown with smaller error bars, whereas larger bars include both statistical and systematic uncertainties.
	}
	\label{Fig4}
\end{figure}

The production fraction ratio $\fcratio$ and the $\Bcm-\Bcp$ production asymmetry are measured by the $\lhcb$ experiment~\cite{LHCb-PAPER-2019-033}.
The analysis is done using data samples of 7~and 13~$\tev$~$\pp$ collisions corresponding to an integrated luminosity of $1.0$ and $1.6~\invfb$, respectively.
The $\B$~mesons in the analysis are reconstructed using the $\BcmToJpsiMumNumb$, $\BdToDpXMumNumb$ and $\BudToDzXMumNumb$ decay modes.
The $\X$~symbol is used here and throughout to refer to any additional undetected particles.
The production fractions are obtained as a function of $\pt$ and $\eta$ of the $\B$~mesons.
The measured production fraction for the 7~$\tev$ data sample is shown in Fig.~\ref{Fig4}\,(a) and~(b).
There is a small dependence on the transverse momentum, but no dependence on $\eta$ is observed.
The ratio of production fractions averaged over $\pt$ and $\eta$ of the $\B$~mesons are measured to be
\vspace*{-2mm}
\begin{align}
\fcratio&=(3.63\pm0.08\pm0.12\pm0.86)\times 10^{-3} \rm \;\, for\;\, 7~\tev,\\
\fcratio&=(3.78\pm0.04\pm0.15\pm0.89)\times 10^{-3} \rm \;\, for\;\, 13~\tev,
\end{align}
\noindent where the first uncertainty is statistical, the second one is systematic and third one is due to the spread of theoretical predictions of the $\BcmToJpsiMumNumb$ branching fraction available in literature (see references in Ref.~\cite{LHCb-PAPER-2019-033}).
Recently the HPQCD collaboration provided the first lattice prediction of the $\BcmToJpsiMumNumb$ decay width, which has an uncertainty of only 10\%~\cite{Harrison:2020gvo}.

The $\Bcm-\Bcp$~production asymmetry is measured in different intervals of \pt and $\eta$ of the $\B$~mesons, no significant asymmetry is observed.
The $\Bcm$~meson production asymmetries, averaged over $\pt$ and $\eta$, are measured to be $(-2.5 \pm 2.1 \pm 0.5)\%$ and $(-0.5 \pm 1.1 \pm 0.4)\%$ for the 7~and 13~$\tev$ data samples, respectively.

\section{Conclusion}
A significant contribution to the knowledge of beauty hadron spectroscopy is provided by the LHCb experiment.
In particular,
the new excited $\Obm$ and $\Lb$~states are observed, mass and width of newly observed states are measured.
The production fraction of $\Bcm$~meson with respect to $\Bu$~and $\Bd$~mesons and $\Bcm-\Bcp$ production asymmetry are measured in $\pp$~collisions at centre-of mass energies of 7~and 13~$\tev$.

\section{Acknowledgements}
This work is supported by the Russian Foundation for Basic Research under grant \mbox{№~20-32-90166}.
The author would like to express his gratitude to the Olga Igonkina Foundation for supporting this talk.
Also, the author would like to express special thanks to the ICHEP'2020 organizers for the great virtual conference.

\bibliographystyle{JHEP}
\bibliography{standard}
\end{document}